\documentclass[preprint,12pt]{elsarticle}

\usepackage{graphicx}

\usepackage{amssymb}

\newcommand{\beq}{\begin{equation}}
\newcommand{\eeq}[1]{\label{#1}\end{equation}}
\newcommand{\beqa}{\begin{eqnarray}}
\newcommand{\eeqa}[1]{\label{#1}\end{eqnarray}}
\newcommand{\eeqan}{\end{eqnarray}}
\newcommand{\CR}{\nonumber \\ }

\journal{Nuclear Physics A}

\begin{document}

\begin{frontmatter}



\title{Chirally motivated $\bar{K}N$ amplitudes\\for in-medium applications}

\author[UJF]{A.~Ciepl\'{y}}
\author[UTEF]{J.~Smejkal}
\address[UJF]{Nuclear Physics Institute, 250 68 \v{R}e\v{z}, Czech Republic}
\address[UTEF]{Institute of Experimental and Applied Physics, Czech Technical University in Prague,\\
Horsk\'{a} 3a/22, 128~00~Praha~2, Czech Republic}

\begin{abstract}
A new fit of a chirally motivated coupled-channels model for meson-baryon interactions is 
presented including the recent SIDDHARTA data on the 1s level characteristics of kaonic hydrogen. 
The kaon-nucleon amplitudes generated by the model are fully consistent with our earlier studies. 
We argue that a sharp increase of the real part of the in-medium $K^{-}p$ amplitude at subthreshold 
energies provides a link between the shallow $\bar{K}$-nuclear optical potentials obtained 
microscopically from threshold $\bar{K}N$ interactions and the phenomenological deep ones deduced 
from kaonic atoms data. The impact on the $A$-dependence of the $\Lambda$-hypernuclear formation 
rates measured in reactions with stopped kaons is discussed too.
\end{abstract}


\end{frontmatter}


\section{Introduction}
\label{sec:intro}

The interaction of kaons with nuclear medium is standardly described in terms 
of kaon-nuclear optical potential constructed as a coherent sum of kaon interactions 
with individual nucleons, $V_{K}(\rho) \sim F_{KN}\, \rho$, where $F_{KN}$ is the effective
elementary kaon-nucleon scattering amplitude and $\rho$ stands for the nuclear density.
For many years the strength of the attractive $K^-$-nuclear optical potential 
remained a puzzle with two conflicting scenarios. The density dependent 
formulation of the $F_{K^{-}N}$ amplitude implemented in phenomenological fits 
to kaonic atoms data gave a strong evidence of a deep $K^{-}$-nuclear 
potential \cite{1994FGB}, in a range ${\rm Re}\:V_{K^-}(\rho_0)\sim -$(150--200) MeV 
at nuclear density $\rho_{0}=0.17~{\rm fm}^{-3}$. On the other hand, models
based on a threshold value of the $\bar{K}N$ amplitude calculated within coupled-channels 
$\pi\Sigma$--$\bar{K}N$ framework with chiral dynamics \cite{1995KSW}, \cite{2003JOORM}, 
\cite{2005BNW}, \cite{2007CS} provide an optical potential which is only (40--50)~MeV deep 
when nuclear medium effects (Pauli blocking and kaon self-energy) are taken into account 
\cite{2000RO}, \cite{2001CFGM}. In a recent work \cite{2011CFGGM} its authors pointed to 
a strong energy dependence of the chirally motivated in-medium $K^{-}p$ amplitude 
and suggested that the puzzling discrepancy could be resolved by evaluating 
the amplitude at subthreshold energies rather than at the $\bar{K}N$ threshold. 

In the present paper we demonstrate that the subthreshold energy dependence of the $K^{-}p$ 
interaction depends only moderately on particularities of the underlying chiral model. 
Specifically, we show that our new fits of model parameters to the fresh experimental data
on kaonic hydrogen characteristics \cite{2011SIDD} generate $\bar{K}N$ amplitudes that are 
quite consistent with those obtained with a model \cite{2010CS} fitted 
to the older kaonic hydrogen data \cite{2005DEAR}. We also present the $\pi\Sigma$ mass 
spectrum generated by the model that appears consistent with observations of the $\Lambda(1405)$ 
resonance peaking at energies around $1400$ MeV, rather then around $1420$ as advocated 
by some other authors \cite{2005MOR}, \cite{2008HW}. Finally, we briefly review the impact 
of energy dependence implemented in evaluation of $K^{-}N$ branching ratios on the calculated 
$\Lambda$-hypernuclear formation rates in the ($K^{-}_{\rm stop}$,$\pi^{-}$) reactions 
\cite{2011FINUDA}, \cite{2011CFGK}.

\section{Separable potentials model}
\label{sec:model}

The synergy of chiral perturbation theory and coupled channel $T$-matrix 
re-summation techniques provides successful description of $\bar{K}N$ 
interactions at low energies. In our approach we employ 
chirally motivated coupled-channel $s$-wave potentials that are taken in 
a separable form, 
\begin{equation}  
V_{ij}(p,p';\sqrt{s})=\sqrt{\frac{1}{2\omega_i}\frac{M_i}{E_i}}\; g_{i}(p)
             \; \frac{C_{ij}(\sqrt{s})}{f_{i}\,f_{j}} \;
             g_{j}(p') \sqrt{\frac{1}{2\omega_j}\frac{M_j}{E_j}} \;\; ,
\label{eq:Vpot} 
\end{equation} 
with $\sqrt{s}$, $p$ and $p'$ denoting the total meson-baryon center-of-mass energy, initial 
and final state meson momenta, respectively. Further, $E_i$, $M_i$ and $\omega_i$ stand for 
the baryon energy, baryon mass and meson energy in the c.m. system of channel $i$. 
The coupling matrix $C_{ij}$ is determined by chiral SU(3) symmetry. The parameters 
$f_{i}$ represent the pseudo-scalar meson decay constants and the Yamaguchi type 
form factors $g_{i}(p)=1/[1+(p/ \alpha_{i})^2]$ are determined by inverse 
range parameters $\alpha_{i}$. The indices $i$ and $j$ run over the meson-baryon 
coupled channels $\pi\Lambda$, $\pi\Sigma$, $\bar{K}N$, $\eta\Lambda$, 
$\eta\Sigma$ and $K\Xi$, including all their appropriate charge states. 
Details of the model are given in Ref.~\cite{2010CS} where a common 
value of $f_{i} = f \sim 100$ MeV was adopted in all channels
and used as a free parameter in fits to experimental data. 
Here we follow the example of Refs.~\cite{2001NR} and \cite{2011IHW} and allow 
for three different couplings $f_{\pi}$, $f_{K}$ and $f_{\eta}$.

The chiral symmetry of meson-baryon interactions is reflected in the structure 
of the $C_{ij}$ coefficients derived directly from the Lagrangian. In practice, 
one often considers only the leading order Tomozawa-Weinberg (TW) interaction 
with energy dependence given by 
\begin{equation}  
C_{ij}(\sqrt{s}) = - C_{ij}^{\rm TW} (2\sqrt{s} -M_{i} -M_{j})/4  \; ,
\label{eq:CTW} 
\end{equation} 
with $C_{ij}^{\rm TW}$ standing for the SU(3) Clebsh-Gordan coefficients. 
The exact content of the matrix elements up to second order in the meson c.m. kinetic 
energies was already specified in Ref.~\cite{1995KSW} and followed in Ref.~\cite{2010CS}. 
In the present work we adopt an alternate formulation of the next-to-leading order (NLO) 
terms for the s-type (direct) and u-type (crossed) Born amplitudes. 
Following the prescription for these two terms by N.~Fettes \cite{2000Fet}, \cite{2000FMMS} 
and keeping only the contributions up to second order in meson momenta the two NLO terms read
\beqa
C_{i j}^{(s)}(\sqrt{s}) & = & 0
\CR
C_{i j}^{(u)}(\sqrt{s}) & = & C_{i j}^{(u)} \frac{1}{M_{0}} \,
        \left( - p_{i}^{2} - p_{j}^{2}
            + \frac{1}{3}\frac{p_{i}^{2}\, p_{j}^{2}}{m_{i}\, m_{j}} \right)\;,
\eeqa{eq:Csu}
where the coefficients $C_{i j}^{(u)}$ are the same as those specified in the Table 11 
of Ref.~\cite{2010CS}, $M_{0}$ stands for the baryon mass in the chiral limit and $m_i$  
denote the meson masses. The main advantages of the adopted formulation are a scalar character 
of the baryon propagator and its angular independence in case of the u-term.
The vanishing of the s-term appears due to averaging over angles in the $s$-wave 
and would not occur if our model involved higher angular momenta. 
The resulting form of our u-term requires additional comments. In a proper NLO expansion, 
there should be a product of meson energies $\omega_{i}$ and $\omega_{j}$ in the denominator 
of the last contribution instead of the masses $m_{i}$ and $m_j$. However, this would make 
the u-term divergent for very deep subthreshold energies where 
$p_{i}^{2} \sim -m_{i}^{2}$. This would lead to unphysical divergences 
in our meson-baryon interaction potential (\ref{eq:Vpot}) that should be cured 
by higher than NLO terms. Since we terminate the chiral expansion 
at the second order we have to regularize the u-term in some other way which 
we do by approximating the meson energies there by their masses. We have checked 
that this approximation is perfectly justified in a large energy interval around 
the $\bar{K}N$ threshold, without any observable impact on the resultinlg $\bar{K}N$ 
amplitudes as far as to $\pi\Sigma$ threshold when extrapolating to $\bar{K}N$ 
subthreshold energies.

When the separable potentials (\ref{eq:Vpot}) are used in coupled channel
Lippman-Schwinger equation the resulting scattering amplitudes are also 
of a separable form given explicitly by 
\begin{equation} 
F_{ij}(p,p';\sqrt{s})=-\frac{g_{i}(p)g_{j}(p')}{4\pi f_{i}\,f_{j}} 
\sqrt{\frac{M_{i}M_{j}}{s}}
\left[(1-C(\sqrt{s})\cdot G(\sqrt{s}))^{-1} \cdot C(\sqrt{s})\right]_{ij}\;. 
\label{eq:ampl} 
\end{equation} 
Here the meson-baryon propagator $G(\sqrt{s})$ is diagonal in the channel 
indices $i$ and $j$. When the elementary $\bar{K}N$ system is submerged in 
the nuclear medium one has to consider Pauli blocking and self-energies (SE)
generated by the interactions of mesons and baryons with the medium. 
Thus, the propagator $G(\sqrt{s})$ and the amplitudes $F_{ij}$ become 
dependent on the nuclear density $\rho$. The intermediate state Green's 
function is calculated as 
\begin{equation} 
G_{i}(\sqrt{s};\rho)=\frac{1}{f_{i}^{2}}\frac{M_i}{\sqrt{s}}\int_{\Omega_{i}
(\rho)}\frac{d^{3}{\vec p}}{(2\pi)^{3}}\frac{g_{i}^{2}(p)}{p_{i}^{2}-p^{2}
-\Pi_{i}(\sqrt{s},\vec{p};\rho)+{\rm i}0}\;\; , 
\label{eq:Green} 
\end{equation} 
where $\vec{p_i}$ is the on-shell c.m. momentum in channel $i$ and the 
integration domain $\Omega_{i}(\rho)$ is limited by the Pauli principle in 
the $\bar{K}N$ channels. Included in the denominator of the Green's function 
(\ref{eq:Green}) is the sum $\Pi_{i}$ of meson and baryon self-energies in 
channel $i$. Since the kaon SE is constructed from the resulting  
$\bar{K}N$  amplitudes a selfconsistent procedure is required 
as first suggested by Lutz \cite{1998Lut}. In our calculation, 
following Ref.~\cite{2001CFGM}, the baryon and pion self-energies 
were approximated by momentum independent potentials $V=V_{0}\:\rho/\rho_{0}$ 
with real and imaginary parts of $V_{0}$ chosen consistently from mean-field 
potentials used in nuclear structure calculations and in scattering 
calculations, respectively. Specifically, we adopted 
$V_{0}^{\pi}=(30-\rm{i}10)$ MeV, $V_{0}^{\Lambda}=(-30-\rm{i}10)$ MeV, 
$V_{0}^{\Sigma}=(30-\rm{i}10)$ MeV and $V_{0}^{N}=(-60-\rm{i}10)$ MeV.

\subsection{Fits to experimental data}
\label{sec:vacuum}

The free parameters of the separable-interaction chiral models considered 
in Ref.~\cite{2010CS} and in the present work were fitted to the available 
experimental data on low energy $\bar{K}N$ interactions consisting of
\begin{itemize}
\item the $K^- p$ cross sections for the elastic scattering and reactions 
(see references collected in \cite{1995KSW}); following the procedure 
adopted in \cite{2010CS} we consider only the data points at  
the kaon laboratory momenta $p_{LAB} = 110$ MeV (for the $K^- p$, 
$\bar{K^0}n$, $\pi^{+} \Sigma^{-}$, $\pi^{-} \Sigma^{+}$ final states) 
and at $p_{LAB} = 200$ MeV (for the same four channels plus $\pi^0 \Lambda$ 
and $\pi^{0} \Sigma^{0}$) 
\item the $K^- p$ threshold branching ratios, standardly denoted as $\gamma$, 
$R_c$, and $R_n$ \cite{1981Mar}
\item the kaonic hydrogen characteristics, the strong interaction shift 
of the $1s$ energy level $\Delta E_{1s}$ and the decay width of the $1s$ 
level $\Gamma_{1s}$ provided by the recent SIDDHARTA measurement \cite{2011SIDD};
the older DEAR data \cite{2005DEAR} were used in Ref.~\cite{2010CS}
\end{itemize}

In general, the chirally motivated models have no problem with reproduction 
of the low energy $K^- p$ cross sections, mostly due to relatively large 
error bars of the experimental data. This goes in line with inclusion 
of the data taken only at two representative kaon momenta to fix 
the cross section magnitude, which turns out sufficient for a good reproduction
of the experimental cross sections in the whole low energy region \cite{2010CS}. 
The threshold branching ratios are determined with much better precision 
and provide a sterner test for any quantitative usage of the models. 
Another stringent test is provided by the new SIDDHARTA measurement \cite{2011SIDD} 
of the kaonic hydrogen data which appears consistent with the other older data 
\cite{1981Mar} on the $K^- p$ reactions at threshold and low energies.

In the current work we present results of two fits to the experimental data, 
one performed with only the leading order (LO) Tomozawa-Weinberg interaction 
(\ref{eq:CTW}) which we denote TW1, and another one that includes the other 
first order plus second order corrections as well and will be referred to as NLO30. 
Since we fit only 15 experimental data 
(10 cross sections, 3 branching ratios, 2 kaonic hydrogen characteristics) 
it is essential to reduce the number of free parameters as much as possible. 
Our TW1 fit was obtained by varying only one inverse range parameter common 
to all channels, $\alpha_i = \alpha_{TW} = 701$ MeV, and one meson decay constant, 
$f_{i} = f_{TW} = 113$ MeV. For the NLO30 fit we fixed the couplings $f_i$ at their 
physical values $f_{\pi} = 92.4$ MeV,  $f_{K} = 110.0$ and $f_{\eta} =118.8$ MeV 
(see references cited in \cite{2011IHW}), and the inverse ranges of the channels 
closed at the $\bar{K}N$ threshold were set to 
$\alpha_{\eta \Lambda} = \alpha_{\eta \Sigma^{0}} = \alpha_{K \Xi} = 700$ MeV. 
The remaining three inverse ranges 
$\alpha_{\pi\Lambda}$, $\alpha_{\pi\Sigma}$ and $\alpha_{\bar{K}N}$ 
were varied together with four low energy constants (denoted as $d$'s, 
see Ref.~\cite{2010CS} for their specification) of the second order chiral Lagrangian. 
The remaining low energy constants (couplings of the second order chiral Lagrangian, 
parameters $b_{D} = 0.064$ GeV$^{-1}$, $b_F = -0.209$ GeV$^{-1}$) were 
fixed to satisfy the Gell-Mann formulas for baryon mass splittings, to give 
the pion-nucleon sigma term $\sigma_{\pi N} = 30$ MeV (parameter $b_0 = -0.190$ GeV$^{-1}$) and 
to reproduce the semileptonic hyperon decays (parameters $D = 0.80$, $F = 0.46$) \cite{2010CS}. 
Thus, the NLO30 model has only 7 free parameters fitted to the data while 
10 free parameters were used in our earlier work \cite{2010CS} and 
even more of them were employed in Ref.~\cite{2011IHW}.
Finally, we note that unlike in Ref.~\cite{2010CS} we keep the energy dependence 
of the TW term in the same form as in our TW1 fit, Eq.~(\ref{eq:CTW}). 
This correction as well as a use of empirical PS meson decay constants introduce 
additional chiral symmetry breaking effects that can be viewed as renormalization 
of the pertinent chiral Lagrangian quantities, thus reducing impact of higher orders 
in the chiral expansion. In Ref.~\cite{2011CFGGM} our TW1 model was already discussed 
and used in analysis of kaonic atoms and kaon-nuclear quasi-bound states alongside 
with an older NLO model CS30 taken from Ref.~\cite{2010CS}. 
Since the CS30 model was fitted to the older DEAR data \cite{2005DEAR}
on kaonic hydrogen it is important to check how much are our conclusions sensitive 
to the mentioned modifications of the NLO model and to the new experimental 
data from SIDDHARTA \cite{2011SIDD}. 

\begin{table}[h] 
\caption{$K^-p$ threshold data calculated in several LO and LO+NLO coupled-channel 
chiral models. The columns show the kaonic hydrogen $1s$ level shift 
$\Delta E_{1s}$ and width $\Gamma_{1s}$ (in eV; the values marked by an asterisk 
were derived from the $K^{-}p$ scattering length by means of the modified Deser-Trueman 
formula \cite{2004MRR}), and the $K^-p$ threshold branching 
ratios $\gamma$, $R_c$, $R_n$. The last two columns list the $I=0$ $S$-matrix 
pole positions $z_1,z_2$ (in MeV) related to the $\Lambda(1405)$ resonance. The last two 
lines show the experimental data and their errors.} 
\begin{tabular}{cccccccc} 
 model  & $\Delta E_{1s}$ & $\Gamma_{1s}$ & $\gamma$ & $R_c$ & $R_n$ &    $z_1$     &    $z_2$      \\ \hline 
 TW1                   & 323       & 659       & 2.36 & 0.636 & 0.183 & (1371,$-$54) & (1433,$-$25)  \\ 
 TW2 \cite{2003JOORM} & 275$^{*}$ & 586$^{*}$ & 2.30 & 0.618 & 0.257 & (1389,$-$64) & (1427,$-$17)  \\
 TW3 \cite{2011IHW}   & 373$^{*}$ & 495$^{*}$ & 2.36 & 0.66  & 0.20  & (1384,$-$90) & (1422,$-$16)  \\[2mm] 
 NLO30                & 310       & 607       & 2.37 & 0.660 & 0.191 & (1355,$-$86) & (1418,$-$44)  \\ 
 CS30 \cite{2010CS}   & 260       & 692       & 2.37 & 0.655 & 0.188 & (1398,$-$51) & (1441,$-$76)  \\ 
 BNW \cite{2005BNW}   & 236$^{*}$ & 580$^{*}$ & 2.35 & 0.653 & 0.194 & (1408,$-$37) & (1449,$-$106) \\ 
 IHW \cite{2011IHW}   & 306$^{*}$ & 591$^{*}$ & 2.37 & 0.66  & 0.19  & (1381,$-$81) & (1424,$-$26)  \\ \hline 
 exp.                 & 283       & 541       & 2.36 & 0.664 & 0.189 & -- & -- \\ 
 error ($\pm$)        &  42       & 111       & 0.04 & 0.011 & 0.015 & -- & -- \\
\end{tabular} 
\label{tab:models} 
\end{table} 

In Table \ref{tab:models} we show how our TW1 and NLO30 models compare with other 
models in reproduction of the $\bar{K}N$ threshold data. The first three lines 
are reserved for models that implement only the TW interaction, the next four 
lines for representative examples of models that incorporate all LO and NLO terms. 
Only the present and the models from Ref.~\cite{2011IHW} use the new SIDDHARTA 
measurement \cite{2011SIDD} of kaonic hydrogen characteristics while the CS30 \cite{2010CS} 
and BNW \cite{2005BNW} models were fitted to the DEAR data \cite{2005DEAR}. 
The model of Ref.~\cite{2003JOORM} was fitted only to the threshold branching 
ratios and to the shape of the $\Lambda(1405)$ resonance 
in the $\pi\Sigma$ mass spectrum. It can be seen that the threshold data are 
quite well reproduced with the models that are based only on the 
TW interaction. When we incorporate the NLO terms the total $\chi^{2}$ per data 
point drops significantly from $3.6$ to $0.61$ for the TW1 and NLO30 models, respectively. 
The CS30 model gives $\chi^{2}/N = 0.78$ for the data set which includes the new 
SIDDHARTA measurement. When the parameters of the CS30 model were fitted 
to the DEAR data \cite{2005DEAR} the resulting $\chi^{2}/N = 1.3$ was almost 
twice as large. This difference demonstrates that the new kaonic hydrogen data \cite{2011SIDD}
from the SIDDHARTA collaboration are much more consistent with the other low energy data on 
$K^{-}p$ scattering and reactions. The two parameter sets of our CS30 and NLO30 models 
(both of them giving $\sigma_{\pi N}=30$~MeV) are compared in the Table \ref{tab:LEC}. 
As expected, the alternate treatment of the NLO terms and use of empirical couplings $f_{i}$ 
has significant impact on the $d$-couplings while the inverse ranges $\alpha_i$ are 
varied only moderately. We also note that our $d$-couplings are somewhat larger than those 
reported for the IHW model \cite{2011IHW}, though there is no simple relation between 
their and our NLO low energy constants.

\begin{table}
\centering
\caption{The inverse ranges (in MeV) and low energy constants $d_0$, $d_D$, $d_F$ and $d_1$ (in GeV$^{-1}$) 
fitted to the available $K^{-}p$ data at threshold and low energies.}
\begin{tabular}{cccccccc}
model &  $\alpha_{\pi\Lambda}$  & $\alpha_{\pi\Sigma}$ & $\alpha_{\bar{K}N}$ & $d_0$  & $d_D$ & $d_F$    & $d_1$  \\ \hline
 CS30  &        291            &         601         &        639          & $-$0.450 & 0.026 & $-$0.601 &    0.235  \\
 NLO30 &        297            &         491         &        700          & $-$0.812 & 0.288 & $-$0.737 & $-$0.016  \\
\end{tabular}
\label{tab:LEC} 
\end{table}

The separate treatment of the inverse ranges in different channels and a relatively 
small value of $\alpha_{\pi \Lambda}$ in our NLO models may require a comment. 
First of all we note that only a small part of the improvement observed when 
going from the LO to the NLO30 fit is due to the separate treatment of 
the inverse ranges in the NLO30 model. In a complementary NLO fit with $f_i$ 
fixed at physical couplings and only one value of inverse range used in all 
channels and varied together with the NLO $d$-couplings the fit to the data 
gave $\chi^{2}/N = 1.1$ for $\alpha_i = \alpha_{NLO} = 481$ MeV. Although the quality 
of this fit is quite good in general it is lacking in respect to the resulting 
kaonic hydrogen $1s$-level width $\Gamma_{1s} = 763$ eV, which is by about two 
standard deviations larger than the SIDDHARTA value (the energy shift 
$\Delta E_{1s} = 275$ eV would be fine). The independent variation 
of inverse ranges in channels open at the $\bar{K}N$ threshold allows to remedy 
this nuisance. It is also a common practice in other $\bar{K}N$ chiral models 
\cite{2003JOORM}, \cite{2005BNW}, \cite{2011IHW} to use channel dependent 
subtraction constants to regularize the intermediate state integral. The large 
negative values of the $\pi \Lambda$ subtraction constant reported in 
the NLO fits of Refs.~\cite{2005BNW} and \cite{2011IHW} are equivalent 
to a very small cutoff momenta \cite{2001OM} which in turn implies small values 
of inverse ranges when the intermediate state integral is regularized by means 
of the Yamaguchi formfactors. Thus, although the relation of our inverse ranges 
to the subtraction constants is highly nontrivial, the small value of 
$\alpha_{\pi \Lambda}$ appears in qualitative agreement 
with the values of the pertinent subtraction constant obtained in other 
NLO fits of the $K^{-}p$ data. 

Finally, we would like to emphasize the importance of calculating the kaonic hydrogen characteristics 
directly by solving the kaon-proton bound state problem rather then deriving them from the $K^{-}p$ 
elastic amplitude extrapolated to the threshold. Standardly, the kaonic hydrogen shift $\Delta E_{1s}$ 
and width $\Gamma_{1s}$ are derived from the $K^{-}p$ scattering length $a_{K^{-}p}$ by means of 
a modified Deser-Trueman formula \cite{2004MRR} taken to the second order in $a_{K^{-}p}$. 
However, we have shown \cite{2007CS} that the precision of the relation is on a level of $10$\%. 
Since the experimental precision of kaonic hydrogen characteristics is getting close to this level 
one should either expand the relation reported in Ref.~\cite{2004MRR} to higher orders of $a_{K^{-}p}$ 
or solve the $K^{-}p$ bound state problem numerically as we do here and in Refs.~\cite{2007CS} and 
\cite{2010CS}.

Also listed in Table \ref{tab:models} are the positions $z_1,z_2$ of the two $I=0$ $S$-matrix poles 
that reside on the second Riemann sheet [$-,+$] of the complex energy manifold, 
where the signs are those of the imaginary parts of the c.m. momenta 
in the $\pi\Sigma$ and $\bar{K}N$ channels, respectively. It is remarkable that all 
the TW models listed in the table are in a relatively close agreement on the position of the 
upper pole $z_2$. This agreement is spoiled, at least in our model (unlike in Ref.~\cite{2011IHW}), 
when the NLO corrections are included in the inter-channel couplings. In contrast, the position 
of the lower pole $z_1$ exhibits model dependence already in the TW models. 
Generally, it is located much further away from the real axis than the pole 
$z_2$. The pole $z_2$ is usually relegated to the subthreshold behavior of 
the $K^{-}p$ amplitude and to the $\Lambda(1405)$ resonance observed in the 
$\pi\Sigma$ mass spectrum in $\bar{K}N$ initiated reactions. 

\begin{figure}[h]    
\centering
\includegraphics[width=0.6\textwidth]{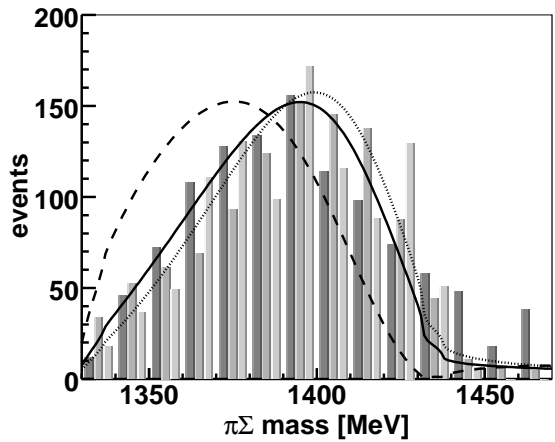}
\caption{The $\pi \Sigma$ mass distribution. Our results are compared with 
the experimental data taken from Refs.~\cite{1973Tho}, \cite{1984Hem} and 
\cite{2008ANKE} with the experimental bars at each energy shown in this order. 
See the text for explanation on the theoretical curves obtained with the NLO30 model.}
\label{fig:pisig}
\end{figure}

The Figure~\ref{fig:pisig} visualizes the $\pi \Sigma$ mass distribution 
computed for the NLO30 model. The solid curve was obtained by tuning the relative 
couplings of the $\pi\Sigma$ and the $\bar{K}N$ channels to an $I=0$ source 
to get a peak at $1395$ MeV, see also Refs.~\cite{2001OM} and \cite{2010CS} for details. Just for 
a reference we also show the spectra obtained by assuming that the 
$I=0$ resonance originates only from the $\pi \Sigma$ channels 
(dashed line in Fig.~\ref{fig:pisig}) or that it is formed exclusively from 
the $\bar{K}N$ channels (dotted line). These two lines represent 
a kind of boundaries on the shape and peak position of the spectra 
in a situation when the low energy constants are fixed at the values 
obtained in our NLO30 fit.  
The experimental data shown in Figure~\ref{fig:pisig} come from three different 
measurements \cite{1973Tho}, \cite{1984Hem}, \cite{2008ANKE}, 
all exhibiting a prominent structure around 1400 MeV. 
As the observed spectra are not normalized we have rescaled the original 
data as well as our computed distributions to give 1000 events in the chosen 
energy interval (from 1330 to 1440 MeV). The three measurements give  
$\pi \Sigma$ distributions that look mutually compatible.
However, as the experimental spectra contain admixtures of $I=1$ and $I=2$ 
contributions they cannot be compared simply with our theoretical predictions 
based on an idea of the $I=0$ source. In reality, a meaningful comparison 
of the measured $\pi \Sigma$ spectra with theory would also require a proper 
treatment of the dynamics and kinematics of the particular reaction, 
i.e. an application of an appropriate reaction model. Since this is out 
of scope of the present work the comparison of the theoretical and experimental 
data is presented in Figure~\ref{fig:pisig} only for illustration.
We also mention that the $K^{-}p \rightarrow \Sigma^{0} \pi^{0} \pi^{0}$ data measured 
by the Crystal Ball Collaboration \cite{2004Pra}, which are not shown in the figure, 
yield a slightly different distribution with a peak structure around 1420 MeV. 
The two identical pions in the final state of this reaction complicate a direct 
comparison with the other experiments, so we also find it questionable to relate 
our computed line-shape to the one observed in \cite{2004Pra}. 

\subsection{Free space and in-medium $\bar{K}N$ amplitudes}
\label{sec:medium}

\begin{figure}[htb]
\centering 
\includegraphics[width=0.95\textwidth]{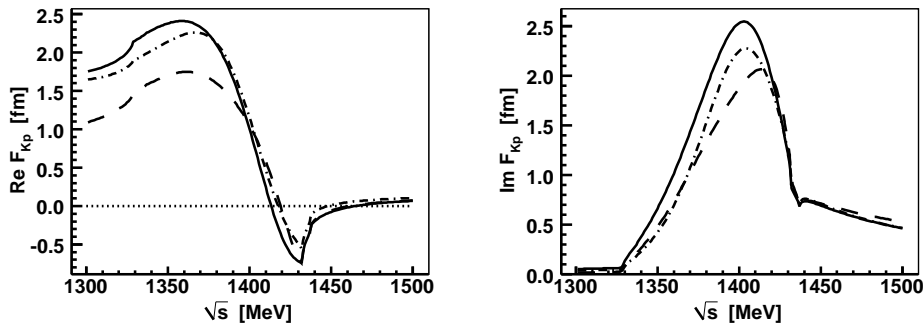}
\caption{Energy dependence of the real (left panel) and imaginary (right panel) parts of 
the elastic $K^-p$ amplitude in the free space. Dashed curves: TW1 model, dot-dashed: CS30 model, 
solid curves: NLO30 model.}
\label{fig:aKp_rho0}       
\end{figure}

The energy dependence of the $K^{-}N$ amplitudes in vacuum and in nuclear medium 
was already discussed extensively in Ref.~\cite{2011CFGGM} where the reduced amplitudes 
(stripped off the form factors $g_i$) were presented. Here we prefer to show 
the full on-shell amplitude $F_{\bar{K}N}$ and anticipate purely imaginary momenta 
$p_{\bar{K}N} = {\rm i}\! \mid \!p_{\bar{K}N}\! \mid$ for the subthreshold energies.
In the present 
Figure \ref{fig:aKp_rho0} we compare the energy dependence of the elastic $K^{-}p$
amplitude in the free space as generated by three different models. The pronounced 
peak in Im~$F_{K^-p}$ and the change of sign in Re~$F_{K^-p}$ point to the existence 
of a quasi-bound state related to the $\Lambda(1405)$ resonance closely below 
the $K^{-}p$ threshold. This feature is common to all models based on chiral dynamics 
which, in combination with couple channel re-summation, generate the resonance 
dynamically. The three models employed in our calculations lead to very similar 
$\bar{K}N$ amplitudes above the threshold and are in qualitative agreement at 
subthreshold energies as well. Interestingly, the Figure \ref{fig:aKp_rho0} also 
demonstrates that the NLO30 model fitted to the SIDDHARTA data \cite{2011SIDD} 
leads to $K^{-}p$ amplitudes that are in close agreement with those obtained 
for the CS30 model fitted to the DEAR data \cite{2005DEAR}. Especially, the real 
parts of the amplitudes generated by the CS30 and NLO30 models are very close 
to each other when extrapolated as far as to (and even below) the $\pi\Sigma$ 
threshold. This feature may be explained by recalling that in fact the CS30 model 
was not able to reproduce the DEAR data and generates kaonic hydrogen characteristics 
that are more compatible with the SIDDHARTA results. In other words, the energy 
dependence of the $K^{-}p$ amplitude is to some extent fixed by the very precise 
threshold branching ratios and by the low energy scattering and reaction data
(see Ref.~\cite{2006BMN} for a detailed analysis). The significance 
of the new SIDDHARTA data can be judged in terms of putting additional 
constrains on the models and reducing the theoretical uncertainties 
when extrapolating the $\bar{K}N$ interaction to subthreshold energies \cite{2011IHW}.

Another point worth mentioning is related to the positions of the poles of the $S$-matrix 
shown in the Table \ref{tab:models}. Standardly, the threshold behavior of the $K^{-}p$ 
amplitude is viewed as strongly affected by the higher of the $I=0$ poles, here labeled as $z_2$. 
We note that this pole is located at rather varied positions for the three different models 
depicted in Figure \ref{fig:aKp_rho0}. For our CS30 model the $z_2$ pole is found even above 
the $\bar{K}N$ threshold and further from the real axis, so it might be the lower $z_1$ pole 
that affects more the subthreshold $K^{-}p$ interaction in this case. Since all our models lead 
to very similar $K^{-}p$ amplitudes at energies at and above the threshold, it looks that 
the positions of the $I=0$ poles cannot be unambiguously determined from the current 
experimental data. In this respect our observation is similar to the one made 
by Shevchenko \cite{2011She} who arrived at a conclusion that the available experimental data 
can be described equally well by models that incorporate either one or two $I=0$ poles.
 
\begin{figure}
\centering 
\includegraphics[width=0.95\textwidth]{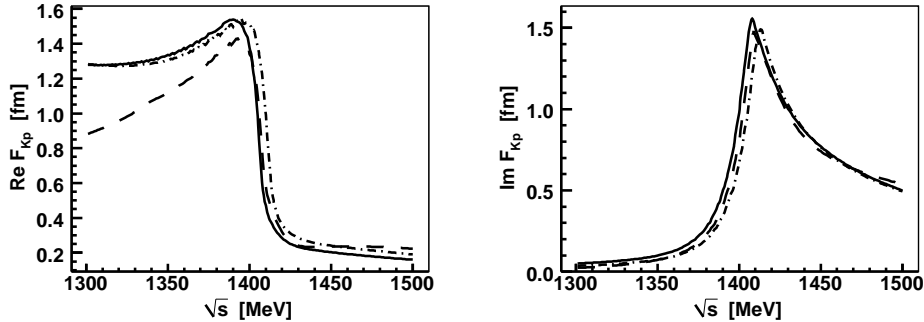}
\caption{Energy dependence of the real (left panel) and imaginary (right panel) parts of 
the elastic $K^-p$ amplitude in nuclear medium at the nuclear density $\rho = \rho_0$. 
Dashed curves: TW1 model, dot-dashed: CS30 model, solid curves: NLO30 model.} 
\label{fig:aKp_rho1}       
\end{figure}

In Figure \ref{fig:aKp_rho1} we show the energy dependence of the elastic $K^{-}p$
amplitude in nuclear medium for the nuclear density $\rho = \rho_0$. Both the Pauli blocking 
as well as the meson and baryon self-energies were included when computing the amplitudes 
in a selfconsistent way. Standardly, only 5--10 iterations are required to achieve 
the selfconsistency \cite{2001CFGM}. The qualitative behavior of the amplitudes is once 
again independent of the model used in the calculations. It is remarkable that 
the selfconsistent treatment leads to even smaller differences between the CS30 and NLO30 
models than they were in the free space, especially in the subthreshold energy region.
However, the prominent subthreshold increase of in-medium $\bar{K}N$ attraction
was not observed in earlier in-medium calculations \cite{2000RO} which get substantially 
different results than ours already when only Pauli blocking is accounted for. 
At the moment we have no real explanation for the difference, maybe the on-shell treatment 
of the intermediate state propagator used in Ref.~\cite{2000RO} is not so well justified 
when the $\bar{K}N$ system is submerged in nuclear medium. On the other hand, we were 
able to reconstruct fully the results of Ref.~\cite{1996WKW} (obtained without accounting 
for the hadron self-energies) when we switched to their parameter set.

As it was already demonstrated in 
Ref.~\cite{2011CFGGM} the in-medium dynamics of the $\Lambda(1405)$ resonance is responsible 
for the rapid increase of the real part of our $K^{-}p$ amplitude at energies about $30$~MeV 
below the $\bar{K}N$ threshold. While the Pauli blocking pushes the resonance structure above 
the threshold the kaon self-energy is responsible for moving it back to energies where it 
is located in the free space. However, when the nuclear density is increased the relevant 
$I=0$ pole crosses the real axis above the $\bar{K}N$ threshold, thus moving to the [$+,-$] 
Rieman sheet. Since it is now located much further from the physical region 
(on a Rieman sheet not connected with the physical one) the $K^{-}p$ amplitude 
no longer exhibits the resonance structure characterized by real part of the amplitude 
crossing zero and imaginary part resembling the Gaussian shape. 

Finally, we mention that the free-space $K^-n$ interaction is weakly attractive and 
its in-medium renormalization is rather weak and exhibits little density dependence 
\cite{2011CFGGM}. We also note that a proper treatment of the $\bar{K}N$ system 
submerged in nuclear medium requires an introduction of realistic in-medium $\bar{K}N$ 
momenta that are used in the Yamaguchi form factors $g_i(p)$ instead of the 
on-shell momenta assumed in Figure \ref{fig:aKp_rho1}. The details can be 
found in Ref.~\cite{2011CFGGM}.

\section{$\Lambda$-hypernuclear production}
\label{sec:hyper}

As mentioned in Section \ref{sec:intro} the energy dependence of the in-medium $\bar{K}N$ 
amplitudes is crucial for a construction of the kaon-nuclear optical potential. In many 
applications the $\bar{K}N$ interaction in nuclear medium is probed at energies below 
the $\bar{K}N$ threshold. The energy shift from threshold to subthreshold energies 
was estimated in Ref.~\cite{2011CFGGM} and an optical potential based on our chirally 
motivated amplitudes was used there to calculate the bound state characteristics 
of kaonic atoms and kaon-nuclear quasi-bound states. The energy dependence of the $K^{-}p$ 
amplitude shown in Fig.~\ref{fig:aKp_rho1} leads to an optical potential that is shallow 
at the $\bar{K}N$ threshold but becomes much deeper at subthreshold energies relevant 
for the $K^{-}$-nuclear bound state systems. Thus, the shift to subthreshold energies 
brings the constructed optical potential in agreement with 
phenomenological analysis of kaonic atoms data that favors deep optical potential \cite{1994FGB}.
In this section we demonstrate another effect of the in-medium energy dependence 
that stems from the energy dependence of the $K^{-}N$ branching ratios.

The new experimental data \cite{2011FINUDA} on the $\Lambda$-hypernuclear production 
in ($K^{-}_{\rm stop}$,$\pi^{-}$) reactions allow to study the $A$-dependence of the formation 
rates for the $p$-shell nuclear targets. The calculated capture rates significantly underestimate 
the measured ones, the deeper the $K^-$ potential, the smaller is the capture rate \cite{2010KCG}. 
However, one can look at relative rates and the $A$-dependence of the production rates where 
the impact of both the theoretical ambiguities and the experimental systematic errors should not 
obscure our observations \cite{2011CFGK}. Within a framework of the distorted wave impulse approximation 
(DWIA) the nuclear capture rate per stopped kaon $R_{fi}/K$ can be expressed as a product of three terms,
a kinematic factor, the elementary branching ratio (BR) of the process BR($K^-N\rightarrow\pi\Lambda$), 
and a rate per hyperon $R_{fi}/Y$ \cite{2010KCG}. While the BR has standardly been taken as a constant 
fixed at its threshold value, our work demonstrates the importance of considering the energy and 
density dependence of the BR.

\begin{figure}
\centering
\includegraphics[width=0.95\textwidth]{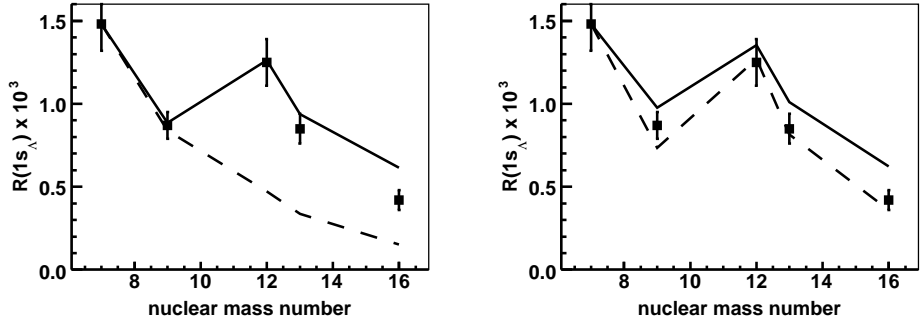}
\caption{The $A$-dependence of the $1s_{\Lambda}$ hypernuclear formation rates, 
experimental data by the FINUDA collaboration \cite{2011FINUDA}. The theoretical 
rates are normalized to the $^{7}$Li experimental value and were calculated with 
a phenomenological density dependent kaon-nuclear optical potential (dashed lines) 
and with the chirally motivated $K^{-}$-nuclear optical potential (solid lines). 
Left panel: elementary BR($K^-N\rightarrow\pi\Lambda$) fixed at the threshold value 
for nuclear density $\rho = \rho_{0}/2$, right panel: energy and density dependent BR.}
\label{fig:rates}       
\end{figure}

In Figure~\ref{fig:rates} we show the calculated $1s_{\Lambda}$ formation rates (per stopped kaon) 
\cite{2011CFGK} in comparison with the FINUDA data \cite{2011FINUDA} for nuclear targets from $^{7}$Li 
to $^{16}$O. The presented theoretical rates were normalized to reproduce exactly the experimental rate 
for the $^{7}$Li target, so one can focus on the $A$-dependence of the rates. The calculated rates in 
the left panel of Fig.~\ref{fig:rates} assume a constant BR generated by the CS30 model, 
taken at the $\bar{K}N$ threshold and for an intermediate nuclear density $\rho = \rho_{0}/2$.
In this case the $A$-dependence is solely driven by the rate per hyperon that contains the overlap 
of the initial and the final state wave functions. There, the chirally motivated kaon-nuclear optical 
potential (based on the CS30 amplitudes) does a better job than the phenomenological one 
(based of the density dependent amplitudes from Ref.~\cite{1994FGB}). The calculated rates presented 
in the right panel of the figure incorporate in addition the energy and density dependence of the BR 
that appears due to an energy shift from threshold to subthreshold $\bar{K}N$ energies. 
Once again we see that the implementation of an energy shift from the $\bar{K}N$ threshold 
to subthreshold energies (where the in-medium $\bar{K}N$ interaction is effective) significantly 
alters the observed picture. 
Both the phenomenological and the chirally motivated optical potentials are sufficiently deep 
at the subthreshold energies relevant for the evaluation of the energy dependent BR, 
so they lead to similar $1s_{\Lambda}$ formation rates.

\section{Summary}
\label{sec:sum}

We have demonstrated that the results of SIDDHARTA experiment \cite{2011SIDD} on kaonic 
hydrogen are in good agreement with other available data on $K^{-}p$ threshold branching 
ratios and on the low energy $K^{-}p$ cross sections. Our new NLO30 model is fully 
compatible with results of our previous analyses and improves the general description 
of the data. 

The several versions of coupled-channels separable potential models 
considered in our work provide $\bar{K}N$ amplitudes that 
exhibit very similar energy dependence in the free space as well as in the nuclear 
medium. Specifically, the strong subthreshold energy and density dependence 
of the $K^{-}p$ amplitudes, that reflects the dominant effect of the $\Lambda(1405)$ 
resonance, does not depend much on a particular version of the model. 
The observed sharp increase of $K^{-}p$ in-medium attraction below the $\bar{K}N$ 
threshold is a robust feature common to all considered models.
It is prominent already in the TW1 model that employs only the leading 
order TW interaction. The NLO contributions present in the CS30 and NLO30 models 
improve significantly the quality of the fit to experimental data but their 
impact on the resulting $\bar{K}N$ amplitudes is only moderate.

Since the $K^{-}$-nuclear interaction probes subthreshold $\bar{K}N$ energies where 
the $K^{-}p$ in-medium amplitude exhibits much stronger attraction the 
resulting $K^{-}$-nuclear optical potential becomes much deeper than 
when it were constructed from the amplitudes taken at the $\bar{K}N$ threshold. 
The mechanism of constructing the optical potential from subthreshold $\bar{K}N$ 
energies allows to link the shallow $\bar{K}$-nuclear potentials based on the chiral 
$\bar{K}N$ amplitude evaluated at threshold and the deep phenomenological optical 
potentials obtained in fits to kaonic atoms data. The relevance of this finding 
to an analysis of kaonic atoms and quasi-bound $\bar{K}$-nuclear states 
was already investigated in Ref.~\cite{2011CFGGM}. The results are also discussed 
in separate reports of this journal \cite{2011FG}, \cite{2011GM}. 

The subthreshold energy and density dependence of the in-medium elementary branching ratio 
BR($K^-N\rightarrow\pi\Lambda$) has a significant impact on the $\Lambda$-hypernuclear 
production rates observed in reactions with stopped kaons. The implementation of the effect 
in our DWIA calculations leads to $A$ dependence of the formation rates that is in very good 
agreement with the one observed in the FINUDA experiment performed on p-shell nuclear targets. 
It also brings in agreement the results obtained with $K^{-}$-nuclear optical potentials 
based on either the chirally motivated or the phenomenological density dependent amplitudes.
Unfortunately, the magnitude of the computed DWIA rates remain much lower than the rates 
established experimentally.

\vspace*{5mm}{\bf Acknowledgement:}
A.~C. acknowledges a fruitful collaboration with E.~Friedman, A.~Gal, D.~Gazda, J.~Mare\v{s}
and V.~Krej\v{c}i\v{r}\'{\i}k who contributed to the papers the current report is partly based on. 
This work was supported by the Grant Agency of the Czech Republic, Grant No. 202/09/1441. 
The work of J.~S. was also supported by the Research Program Fundamental experiments 
in the physics of the micro-world, No. 6840770040, of the Ministry  of Education, Youth 
and Sports of the Czech Republic.




\bibliographystyle{model1a-num-names}
\bibliography{<your-bib-database>}



\end{document}